\documentclass[conference]{IEEEtrans2}
\IEEEoverridecommandlockouts

\usepackage{cite}
\usepackage{amsmath,amssymb,amsfonts}
\usepackage{algorithmic}
\usepackage{bm}
\usepackage{url}
\usepackage{graphicx}
\usepackage{textcomp}
\usepackage{pifont}
\usepackage{tabularx}
\usepackage{amssymb}
\usepackage{multirow}
\usepackage{threeparttable}
\usepackage{makecell}
\usepackage{rotating}
\usepackage{multicol}
\usepackage{booktabs}
\usepackage{hyperref}
\usepackage{xcolor}
\def\BibTeX{{\rm B\kern-.05em{\sc i\kern-.025em b}\kern-.08em
    T\kern-.1667em\lower.7ex\hbox{E}\kern-.125emX}}
\begin{document}

\title{Time-Graph Frequency Representation with Singular Value Decomposition for Neural Speech Enhancement\\
\thanks{This work was supported by the National Natural Science Foundation of China (Grant 62071242).}
}

\author{\IEEEauthorblockN{Tingting Wang$^1$,Tianrui Wang$^2$, $^{\dagger}$ Meng Ge$^2$,  Qiquan Zhang$^3$, Zirui Ge$^1$, Zhen Yang$^1$ \thanks{$^{\dagger}$ Meng Ge is the corresponding author.}}
\IEEEauthorblockA{\textit{$^1$Nanjing University of Posts and Telecommunications, Nanjing, China},\\
\textit{$^2$Tianjin Key Laboratory of Cognitive Computing and Application, College of Intelligence and Computing, }\\
\textit{Tianjin University, Tianjin, China,}
\textit{$^3$The University of New South Wales, Sydney, Australia.}}}


\maketitle
\begin{abstract}
Time-frequency (T-F) domain methods for monaural speech enhancement have benefited from the success of deep learning. Recently, focus has been put on designing two-stream network models to predict amplitude mask and phase separately, or, coupling the amplitude and phase into Cartesian coordinates and constructing real and imaginary pairs. However, most methods suffer from the alignment modeling of amplitude and phase (real and imaginary pairs) in a two-stream network framework, which inevitably incurs performance restrictions. In this paper, we introduce a graph Fourier transform defined with the singular value decomposition (GFT-SVD), resulting in real-valued time-graph representation for neural speech enhancement. This real-valued representation-based GFT-SVD provides an ability to align the modeling of amplitude and phase, leading to avoiding recovering the target speech phase information. Our findings demonstrate the effects of real-valued time-graph representation based on GFT-SVD for neutral speech enhancement. The extensive speech enhancement experiments establish that the combination of GFT-SVD and DNN outperforms the combination of GFT with the eigenvector decomposition (GFT-EVD) and magnitude estimation UNet, and outperforms the short-time Fourier transform (STFT) and DNN, regarding objective intelligibility and perceptual quality. We release our source code at: https://github.com/Wangfighting0015/GFT\_project.
\end{abstract}


\section{Introduction}
Speech enhancement aims to reconstruct the clean speech signal from the noisy speech recording, which often serves as a front-end module for diverse speech applications, such as automatic speech recognition (ASR), speech coding, hearing aids, and speaker recognition~\cite{loizou,overview2018}. In the past decade, with the flourishing of deep learning technology, speech enhancement has shown great progress, showcasing excellent performance in suppressing highly non-stationary noise sources than traditional statistics-based methods~\cite{mmse2017,mmse,zhang2019,8740919}.

Existing neural strategies for speech enhancement mainly utilize either waveform-based~\cite{demcus,kolbaek2020loss} or spectrogram-based techniques~\cite{DeepMMSE,2018complex,yongxu2015}. Due to speech and noise patterns tend to be more distinguishable after the short-time Fourier transform (STFT), the spectrogram-based solutions are still the mainstream solutions, such as UNet~\cite{DBLP:conf/miccai/RonnebergerFB15}, DCCRN~\cite{HuLLXZFWZX20}, TF-GridNet~\cite{wang2023tf}, and the most recent Mamba-based scheme~\cite{zhang2024mamba}. Initially, spectrogram-based methods focused on estimating the magnitude only, and reconstructing the waveform using the estimated magnitude and the original noisy phase. This damaged phase directly limits the upper speech quality of the reconstructed waveform. To alleviate the issue, recent studies have moved to the decoupling-style speech enhancement strategy, including magnitude-phase and real-imaginary decoupling methods. The magnitude-phase decoupling methods often employ a dual-stream network architecture to separately reconstruct the magnitude and phase components of the target speech, while the real-imaginary decoupling methods achieve the reconstruction of the target speech's magnitude and phase indirectly by estimating its real and imaginary components. \textcolor{red}{Additionally, Lu et al.~\cite{LuAL23} propose a dual-path speech enhancement model (MP-SENet), which performs parallel denoising of magnitude and phase spectrum.} Thus, the real-imaginary decoupling methods naturally avoid the phase wrapping issue inherent in magnitude-phase decoupling methods, reconstructing the high-fidelity speech signals.

However, such decoupling-style solutions require the exchange of information between the two streams in the designed network to enhance magnitude and phase estimation. This makes it challenging to ensure the information alignment between the two branch networks. For example, Yin~et\ al.~\cite{DBLP:conf/aaai/YinLXZ20} observes that when employing the real-imaginary decoupling method (e.g., cIRM~\cite{cirm}) for speech enhancement, the real component is accurately estimated, but the imaginary part is almost zero. In other words, the targets of the two branches (i.e., magnitude and phase) are not equally restored, resulting in misalignment. This potential misalignment issue leads to errors during signal reconstruction, resulting in a degradation of speech quality. Thus, this observation prompts a research question: ``How is alignment modeling of magnitude and phase achieved in neural speech enhancement?''.

To answer this question, we first observe that alignment issue arises from the design of dual-stream networks. The key to resolving this issue lies in how to eliminate dual-stream networks while still retaining magnitude and phase information. Motivated by this, we propose the Graph Fourier Transforms (GFTs) using the Singular Value Decomposition (SVD), called GFT-SVD. The proposed GFT-SVD is capable of projecting raw waveform to a real-valued time-graph representation, rather than the complex-valued representation used in traditional STFT, achieving a near-lossless conversion in the process. This process naturally avoids the requirement for dual-stream network modeling of complex-value representations, and retains both magnitude and phase. 

\section{PRELIMINARY WORK}
\subsection{Graph Representation of Speech Signal}
Given a speech frame, the directed graph representation ${\cal{G}}_s$ of speech samples $\bf s$ in this speech frame is denoted as ${{\cal{G}}_s} = \left( {\cal V}_s, {\bf A}_{k}\right)$, 
where ${\cal V}_s$ represents the vertex set and $|{\cal V}_s|=N$. 
${{\rm{A}}_k}(i,j)=1$ if the dependency between the $i_{th}$ speech samples residing on vertex ${v_{i}}$ and the $j_{th}$ speech samples residing on vertex ${v_{j}}$ exists, otherwise ${{\rm{A}}_k}(i,j) =0$~\cite{ref1, ref9}. 
$k$ indicates that there is a dependency between a speech sample and neighboring $k$ samples. 
A graph interpretation of speech graph signals can be achieved by viewing the 0-1 shift matrix as the graph adjacency matrix. 

\subsection{Short-Time Fourier Transform vs. Graph Fourier Transform}
Given the framed speech signal $\textbf s$, the Short-Time Fourier Transform (STFT) can convert the time-domain waveform into a complex-valued frequency-domain spectrogram, defined as follows:
\begin{equation}
{{{{\textbf S}}}}=\text{STFT}{({\textbf s})} = {{\textbf S}_r + j{\textbf S}_i}{\in \mathbb{C}}{\rm{.}} 
\end{equation}
where ${{\textbf S}_r}$ and ${{\textbf S}_i}$ are real and imaginary parts of the complex-valued spectrogram of $\textbf s$.

The Graph Fourier Transform (GFT) is similar to the STFT, consisting of two parts: the Graph Fourier Transform and the inverse Graph Fourier Transform. For example, the Graph Fourier Transform of a signal $\textbf  s$ can be denoted as:
\begin{equation}
{{{{\textbf S}_G}}} = {\text{GFT}{(\textbf s})}={{\bf U^{-1}}}{\textbf s}= {{\textbf S}_G^r + j{\textbf S}_G^i}{\in \mathbb{C}}{\rm{.}} 
\end{equation}
Here, $\bf U$ is the eigenvector matrix of ${\bf A}_{k}$ with the eigenvector decomposition (EVD) method, i.e., ${\bf A}_{k}=\bf U \Lambda U^{-1}$ where ${\bf \Lambda}$ is a diagonal matrix of which diagonal elements $\lambda _0,\lambda _2,...,\lambda _{{N-1}}$ denote the graph frequency. Consider that ${\bf A}_{k}$ is not symmetric matrix, the graph Fourier basis $\bf U$ obtained by the EVD method  is a complex eigenvector matrix according to basic matrix decomposition theory. ${{\textbf S}_G^r}$ and ${{\textbf S}_G^i}$ are the real and imaginary parts of the complex graph spectral of $\textbf s$. Thus, the inverse Graph Fourier Transform is denoted as
\begin{equation}
{{{{\textbf s}}}} = {\text{iGFT}{({{\textbf S}_G}})} = {\bf U}{{\textbf S}_G}{\rm{.}} 
\end{equation}

Both STFT and GFT convert waveforms into complex-valued spectrograms, which are challenging to model directly in networks. As a result, decoupling-style network modeling has become a trend, but this strategy also brings the problem of misalignment between the decoupled elements.


\section{METHODOLOGY}
\label{sec:methodology}
The key difference of our proposed approach lies in the encoder and decoder. Unlike the STFT/GFT based encoders and decoders for complex-valued signal transformation, our GFT-SVD leverages Singular Value Decomposition (SVD) to convert signals losslessly into real-valued spectrograms. This approach avoids the misalignment issue caused by decoupling network modeling, thereby minimizing reconstruction losses.

\vspace{-10pt}
\subsection{Overview}
\begin{figure}[t]
	\centering
	\includegraphics[width=0.7\linewidth]{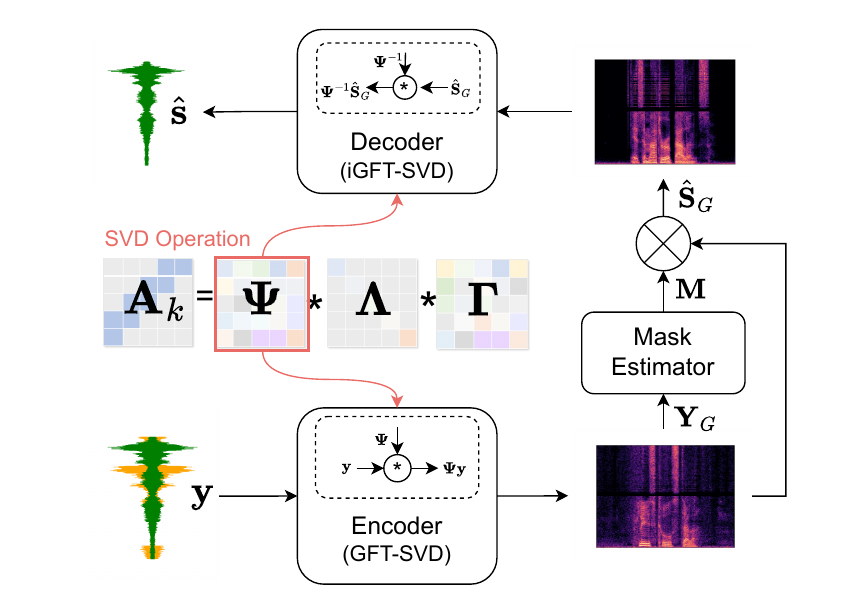}
	\vspace{-5pt}
	\caption{The overview of our neural speech enhancement with GFT-SVD.}
	\label{framework}
	\vspace{-12pt}
\end{figure}

\vspace{-5pt}
\label{subsec:gnn}
Our proposed approach consists of three parts, including encoder, mask estimator, and decoder, as show in Fig.~\ref{framework}. Both the encoder and decoder utilize the designed real-valued Graph Fourier Transform, specifically GFT-SVD and inverse GFT-SVD (iGFT-SVD), respectively. The mask estimator can be implemented using arbitrary suitable enhancement network. 

Formally, given a framed noisy speech $\textbf y$, after passing through the GFT-SVD encoder, it is transformed into a real-valued spectrogram real-valued time-graph representation ${\bf Y}_{\cal {G}}$. Next, the time-graph representation ${\bf Y}_{\cal {G}}$ is fed into the mask estimator to predict the corresponding masks $\bf M$. Finally, the estimated mask is applied to the noisych for denoising, and the clean target speech is reconstructed using the iGFT-SVD. The whole process can be denoted as
\begin{gather}
	\vspace{-1pt}
    \bf{Y}_{\cal{G}} = \text{GFT-SVD}(\textbf{y}) \in \mathbb{R} \label{eq:encoder}\\
    {\bf \hat{S}}_{G}={\bf M} \odot {\bf Y}_{\cal {G}},
    \hat{\textbf{s}} = \text{iGFT-SVD}({\bf \hat{S}}_{G}) \in \mathbb{R}. \label{eq:decoder}
    	\vspace{-1pt}
\end{gather}
As seen above, our pipeline operates entirely with real values, avoiding the requirement for traditional dual-branch decoupling structures used to handle complex-valued spectrogram.


\subsection{Real-Valued Graph Fourier Transform}
\label{subsec:dvc}
Differing to the graph Fourier basis with EVD, we use singular value decomposition (SVD) method that realizes arbitrary shape matrix decomposition to decompose  0-1 real-valued matrix ${\bf A}_{k}$, which uses its unitary matrices to define a new graph Fourier basis. That is,  
${\rm{SVD}}\left( {{\textbf A}_{k}} \right) = {\bf \Psi} \times {\bf \Lambda} \times {{\bf \Gamma}}{\rm{,}}$ where the unitary matrix ${\bf \Psi}$ and ${\bf \Gamma}$ are left and right singular eigen matrix of ${{\bf{A}}_k}$, $\times$ represents the multiply operation.

By utilizing the unitary matrix property, ${\bf \Psi}$ that consists of real-valued eigen vector can map the noisy speech $\textbf y$ into a real-valued graph frequency domain.
Thus, the real-valued Graph Fourier Transform of $\textbf{y}$, i.e., Eq. (\ref{eq:encoder}), is given by 
\begin{equation}
{{{{\textbf Y_G}}}} = \text{GFT-SVD}{({\textbf y})}={{\bf \Psi}}{\textbf y}.  
\end{equation}
Similarly, the inverse graph Fourier transform of $\hat{\textbf{S}}_G$ in Eq. (\ref{eq:decoder}) can be calculated by the inverse of ${\bf \Psi}$, which is denoted as
\begin{equation}
\hat{\textbf{s}} = \text{iGFT-SVD}({\bf \hat{S}}_{G}) ={{\bf \Psi}^{-1}}{\hat{\textbf{S}}_G}{\rm{.}} 
\end{equation}
\begin{figure}[h]
    \begin{minipage}[t]{0.3\linewidth}
        \includegraphics[width=1.3in, height=3.0cm]{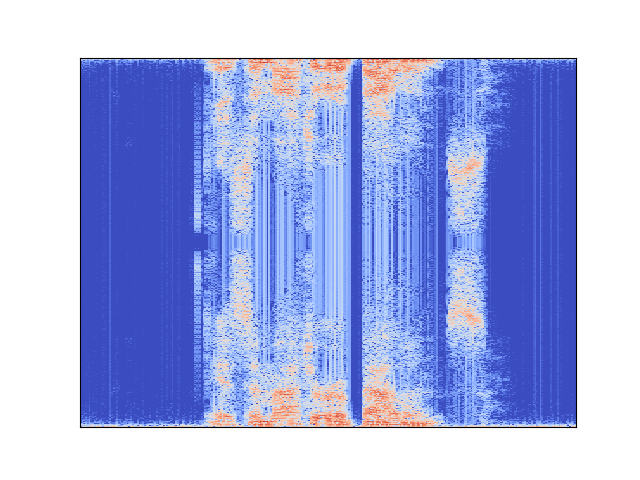}\vspace{-0.4cm}
        \centerline{\scriptsize \ \ \ \ \ \  (a)real-part(STFT) }
    \end{minipage}%
    \begin{minipage}[t]{0.3\linewidth}
    \includegraphics[width=1.3in,height=3.0cm]{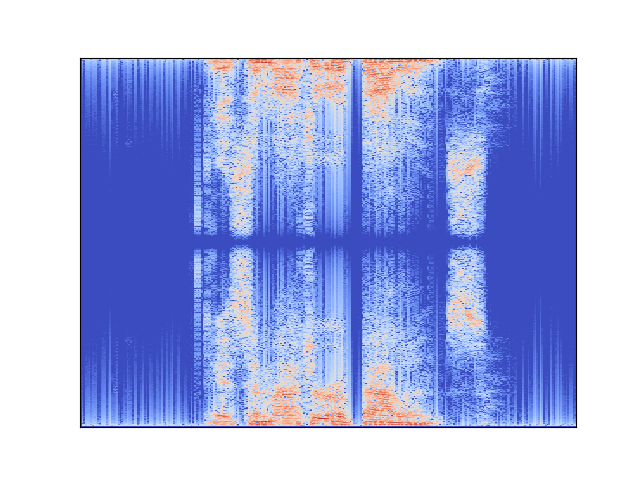}\vspace{-0.4cm}
        \centerline{\scriptsize \ \ \ \ \  (b)imaginary-part(STFT)}  \end{minipage}
 \begin{minipage}[t]{0.3\linewidth}
    \includegraphics[width=1.3in,height=3.0cm]{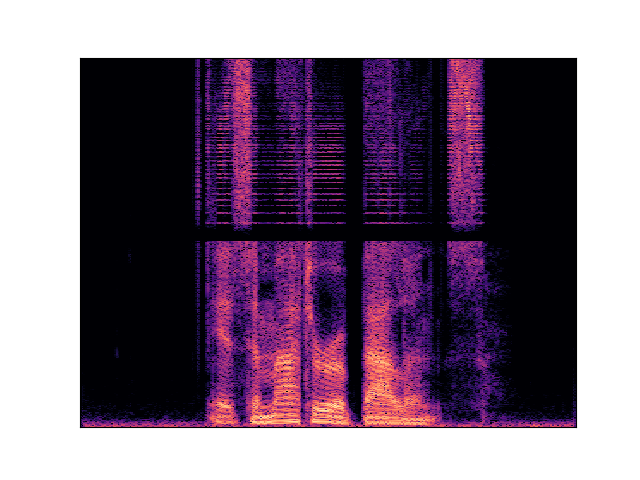}\vspace{-0.4cm}
        \centerline{\scriptsize \ \ \ \ \ \  \ \ \ \  (c)graph spectral (GFT-SVD)}
    \end{minipage}\vspace{-0.25cm}
     \caption{Example of the clean speech. The real and imaginary parts of clean speech after STFT, and the graph spectral of clean speech after GFT-SVD are visualized for illustration convenience.}
    \label{fig:spectrm}
    \vspace{-5pt}
\end{figure}
Fig.~\ref{fig:spectrm} shows the visualization of real and imaginary parts of clean speech after STFT and that of the graph spectra of clean speech after GFT-SVD.  
 Observe from Fig.~\ref{fig:spectrm}, we can see that the real and imaginary parts are symmetrical after STFT extended by traditional symmetrical Fourier series,  while the real-valued graph spectra is unsymmetrical and contains amplitude and phase synchronously. It  differs from the traditional spectrum with STFT by two steps, which helps the mask trains faithfully better.

\subsection{Mask Estimator and Training Objective}
\label{subsec:dvc}
\vspace{-1pt}
As mentioned earlier, our designed GFT-SVD based encoder and decoder can be adapted to any masking-based enhancement network. Here, we use various networks as mask estimator, including NSnet2~\cite{XiaBRDCT20}, dual-path convolution recurrent network (DPCRN)~\cite{abs-2107-05429}, dense convolutional recurrent network (DCRN)~\cite{PandeyLWS21}, deep complex convolution recurrent network (DCCRN)~\cite{HuLLXZFWZX20}, multi-scale temporal frequency convolutional network with axial self-attention (MTFAA)~\cite{ZhangWYW22}, UNet~\cite{DBLP:conf/miccai/RonnebergerFB15}, band-split RNN (BSRNN)~\cite{bsrnn} and Graph UNet (G-UNet)~\cite{ZhangP22} .

 

The training objective of the entire network is based on the SI-SDR loss~\cite{wangPGYL23} between the estimated waveform $\hat{\textbf{s}}$ and the clean speech $\textbf{s}$, which is denoted as follow.
\begin{equation}
    \label{eqa:loss_sisnr}
    \mathcal{L}_{\text{SI-SDR}} = 20 \log_{10} \frac{||\frac{<\hat{\bf{s}},{\textbf{s}}>{\textbf{s}}}{||{\textbf{s}}||^2}||}{||\hat{\textbf{s}} - \frac{<\hat{\textbf{s}},{\textbf{s}}>{\textbf{s}}}{||{\textbf{s}}||^2}||}.
\end{equation}

\vspace{-1pt}
\section{Experimental Results}
\label{sec:setups}
\vspace{-1pt}

\subsection{Dataset and Experimental Setup}
We verify the effectiveness of the proposed approach on two benchmark datasets: DNS-2020 and VCTK+DEMAND~\cite{LvHZX21,DBLP:conf/interspeech/ReddyGCBCDMAABR20}. The recordings of DNS-2020 dataset are from the 2020 Deep Noise Suppression Challenge, and are sampled at 16 kHz. We generate 100-hour noisy data with SNR ranges from -5dB to 20dB using DNS-2020. The training and validation sets are partitioned from the training data at a ratio of 4:1. VCTK+DEMAND is a simulated dataset. The clean speech is selected from 28 speakers in the Voice Bank corpus~\cite{DBLP:conf/ococosda/VeauxYK13}, while the noisy speech is generated by mixing the clean speech with noise from the Diverse Environments Multichannel Acoustic Noise Database (DEMAND)~\cite{DBLP:DEMAN}. The ratio of training to validation utterances in the VCTK+DEMAND dataset ios approximately 9:1.

For experimental setup, the window length and hop size are 25 ms and 6.25 ms for all models. The GFT length is 512. 512-dimension graph features are fed into NSnet2, DPCRN, DCRN, DCCRN, MTFAA, UNet, BSRNN and G-UNet networks. Consider that our graph frequency spectrum by GFT-SVD belongs to the real-valued domain, the band merging and splitting process of backbone MTFAA and BSRNN models are removed. NSnet2, DPCRN, DCRN, DCCRN, UNet and G-UNet apply the originally proposed corresponding network. The 512-dimension features after STFT are inputted into the baseline for fairness of comparisons. 

Perceptual evaluation of speech quality (PESQ) \cite{DBLP:conf/icassp/RixBHH01}, short-time objective intelligibility (STOI)~\cite{DBLP:conf/icassp/TaalHHJ10}, and scale-invariant signal-to-distortion ratio (SI-SDR)~\cite{DBLP:conf/icassp/RouxWEH19} are employed for objective evaluation of enhanced speech signals in the paper.

\begin{table}[t]
	\caption{PERFORMANCE COMPARISON OF DIFFERENT MODELS BASE ON GFT-SVD WITH ${\textbf A}_{3}$ ON THE NO-REVERB DATASET}
    \small
    \def\arraystretch{0.91}
	\setlength{\tabcolsep}{2.7pt}
  \begin{tabular}{lcccccccc}
    \toprule[1.5pt] 
   \bf{Neural}&  \bf Fourier&  \multirow{2}*{\bf{WB-PESQ}}  & \multirow{2}*{\bf{NB-PESQ}}  & \bf{SI-SDR}  & \multirow{2}*{\bf{STOI}}  \\
   \bf{Network}& \bf Transform &  & & (dB) &  
   \\\midrule \midrule 
   	NSnet2~\cite{XiaBRDCT20}& \multirow{1}*{STFT}  & 2.331&2.865& {16.262}&{0.946}\\
     \textbf{NSnet2$_{\cal {G}}$} & GFT-SVD &{2.439}&{3.008}& {16.352}& {0.950}\\ \midrule
		DPCRN~\cite{abs-2107-05429}& STFT & {2.797}&{3.260}& {18.570}& {0.967}\\ 
\textbf{DPCRN$_{\cal {G}}$} & GFT-SVD&{2.885}&{3.392}& {18.964}& {0.968}\\ \midrule
		DCRN~\cite{PandeyLWS21}& STFT & 2.566&{3.074}& {17.326}& {0.958} \\ 
  \textbf{DCRN$_{\cal {G}}$} &GFT-SVD&{2.739} &{3.272}& {18.143 }& {0.962} \\ \midrule
		DCCRN~\cite{HuLLXZFWZX20}& STFT & {2.644}&{ 3.148}& {17.950}&{0.963}\\
  \textbf{DCCRN$_{\cal {G}}$} &GFT-SVD & {2.841}& {3.376}& {18.587}& {0.967}\\\midrule
		MTFAA~\cite{ZhangWYW22}& STFT & {2.696}&{3.243}& {18.294}&{0.964}\\ 
  \textbf{MTFAA$_{\cal {G}}$} & GFT-SVD& 2.707&3.261& 18.454& 0.963\\\midrule
        BSRNN~\cite{bsrnn}& STFT & {2.203}&{2.734}& {16.070}&0.942  \\
        \textbf{BSRNN$_{\cal {G}}$} &GFT-SVD & {2.321}&{2.835}& {16.300}&0.943  \\  \midrule
        \midrule
        UNet~\cite{DBLP:conf/miccai/RonnebergerFB15}  &\multirow{1}*{STFT}& {2.017}&{2.597}& {14.585}&0.940   \\
        G-UNet~\cite{ZhangP22}  &GFT-EVD &2.580 & 3.129&17.583 &0.959 \\       
        {\textbf{UNet}$_{\cal {G}}$} &GFT-SVD &{2.785} &{3.346}&{18.466} &{0.966} \\ 
      \toprule[1.5pt]
  \end{tabular}
  \vspace{-1.5em}
  \label{table_DNSchallengs}
\end{table}

\begin{table}[t]
	\caption{PERFORMANCE COMPARISON OF DIFFERENT MODELS BASE ON GFT-SVD WITH ${{\textbf A}_{3}}$ ON THE VCTK+DEMAND DATASET}
    \small
    \def\arraystretch{0.91}
	\setlength{\tabcolsep}{2.7pt}
  \begin{tabular}{lcccccccc}
    \toprule[1.5pt] 
  \bf{Neural}&  \bf Fourier& \multirow{2}*{\bf{WB-PESQ}}  & \multirow{2}*{\bf{NB-PESQ}}  & \bf{SI-SDR}  & \multirow{2}*{\bf{STOI}}  \\
   \bf{Network}& \bf Transform &  & & (dB) &  
   \\\midrule \midrule 
   	NSnet2~\cite{XiaBRDCT20}& \multirow{1}*{STFT} & 2.343&3.109& {17.793}&{0.928}\\
    	\textbf{NSnet2$_{\cal {G}}$} &GFT-SVD&{2.462} &{3.215}&{18.035} &{0.932}\\  \midrule 
		DPCRN~\cite{abs-2107-05429} &STFT & {2.551}&{3.333}& {18.544}& {0.940}\\ 
  \textbf{DPCRN$_{\cal {G}}$} &GFT-SVD &{2.741}&{3.556}& {18.826}&{0.944}\\ \midrule 
		DCRN~\cite{PandeyLWS21} &STFT & 2.436&{3.211}& {17.877}& {0.933} \\
  \textbf{DCRN$_{\cal {G}}$}&GFT-SVD &{2.624}&{3.426}& {18.242}&{0.937}\\ \midrule 
		DCCRN~\cite{HuLLXZFWZX20} &  STFT& {2.510}&{ 3.255}& {18.242}&{0.933}\\
  \textbf{DCCRN$_{\cal {G}}$}& GFT-SVD & {2.634}& {3.468}& {18.125}&{0.940} \\\midrule 
		MTFAA~\cite{ZhangWYW22} &STFT& {2.657}&{3.478}& {18.837}&{0.938}\\ 
  \textbf{MTFAA$_{\cal {G}}$}& GFT-SVD&2.748 &3.553&18.670 & 0.942\\\midrule 
        {BSRNN}~\cite{bsrnn}&STFT& {2.444}&{3.190}& {18.831}&0.928  \\
        \textbf{BSRNN$_{\cal {G}}$} & GFT-SVD&{2.567}&{3.358}& {19.219}& {0.932}\\ \midrule 
        UNet~\cite{DBLP:conf/miccai/RonnebergerFB15} & \multirow{1}*{STFT} & {2.160}&{3.000}& {16.484}&0.926   \\
         G-UNet~\cite{ZhangP22}&GFT-EVD &2.455 & 3.274&18.009 &0.932 \\ 
       \textbf{UNet$_{\cal {G}}$}& GFT-SVD &{2.644} &{3.490}&{18.611} &{0.937} \\ 
      \toprule[1.5pt]
  \end{tabular}
  \vspace{-1.5em}
  \label{table_Voicebank+Demand}
\end{table}

\subsection{Effectiveness Analysis of GFT-SVD}
Table~\ref{table_DNSchallengs} reports the comparison results of models on the DNS-2020 no-reverb testset, in terms of four metrics. NSnet2$_{\cal {G}}$, DPCRN$_{\cal {G}}$, DCRN$_{\cal {G}}$, DCCRN$_{\cal {G}}$, MTFAA$_{\cal {G}}$, UNet$_{\cal {G}}$ and BSRNN$_{\cal {G}}$ represent the corresponding backbone network with the GFT-SVD. From Table~\ref{table_DNSchallengs}, it can be observed that these models of the time-graph representation provide 0.2 gains of average PESQ compared to the backbone networks with STFT and the combination of GFT-EVD and magnitude estimation UNet. In terms of SI-SDR, DPCRN$_{\cal {G}}$, DCRN$_{\cal {G}}$ and UNet$_{\cal {G}}$ can provide more than 1dB gains over the DPCRN, DCRN, and G-UNet respective. 

Table~\ref{table_Voicebank+Demand} shows the comparison results of models on the small but commonly-used dataset VCTK+DEMAND. So that we can fairly compare the backbone network based on GFT-SVD with the combination of backbone networks and STFT/GFT-EVD. From Table~\ref{table_Voicebank+Demand}, the combination of backbone networks and GFT-SVD has a large gain over the combination of backbone networks like DPCRN, DCRN, and UNet with STFT/GFT-EVD on all the four metrics. This proves the advantage of our time-graph representation based on GFT-SVD for neural speech enhancement. 
\vspace{-3pt}
\subsection{Impact of $k$ Values on GFT-SVD}
\begin{figure}[h]
    \vspace{-0.2cm}
    \begin{minipage}[t]{0.5\linewidth}
        \centering
        \includegraphics[scale=0.13]{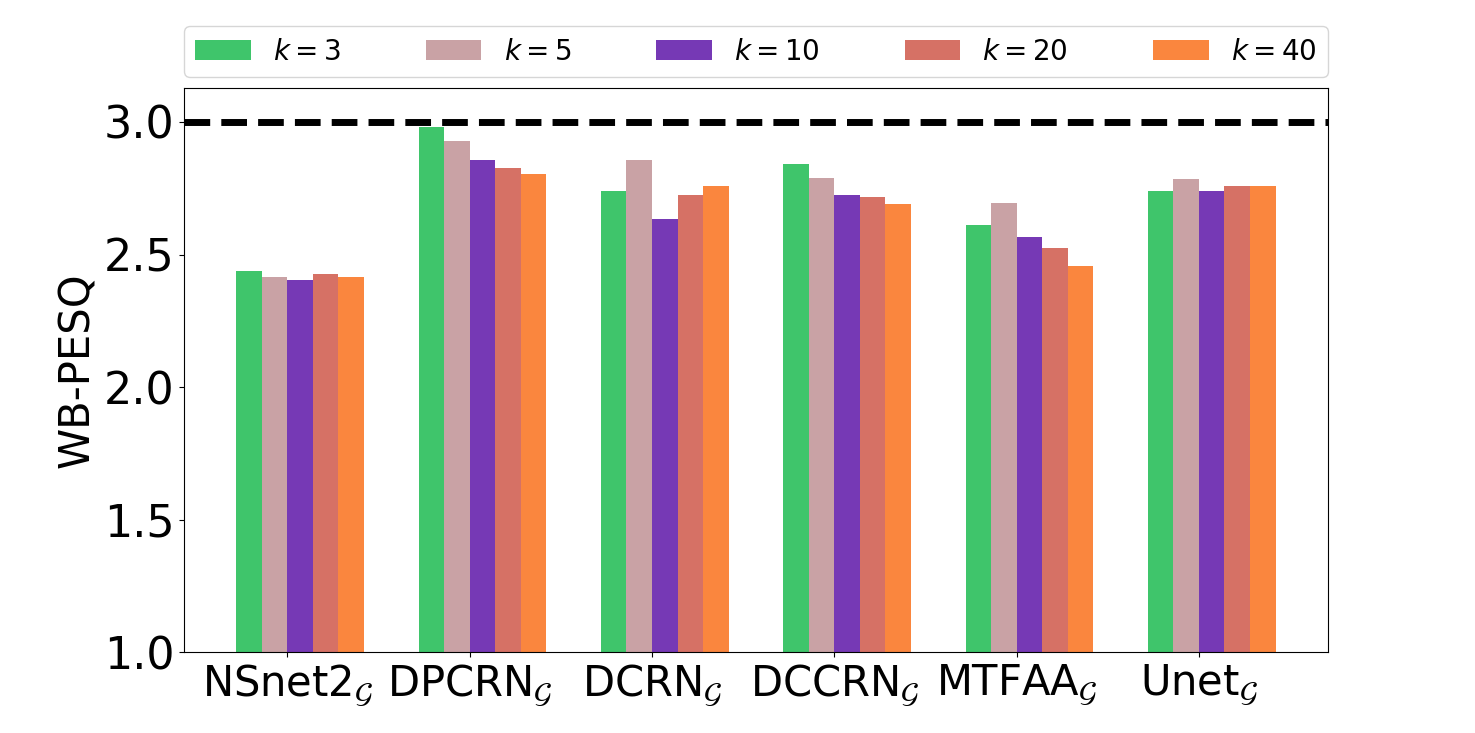}\vspace{-0.2cm}
        \centerline{\scriptsize (a) The average WB-PESQ results }
    \end{minipage}%
    \begin{minipage}[t]{0.5\linewidth}
        \centering
        \includegraphics[scale=0.13]{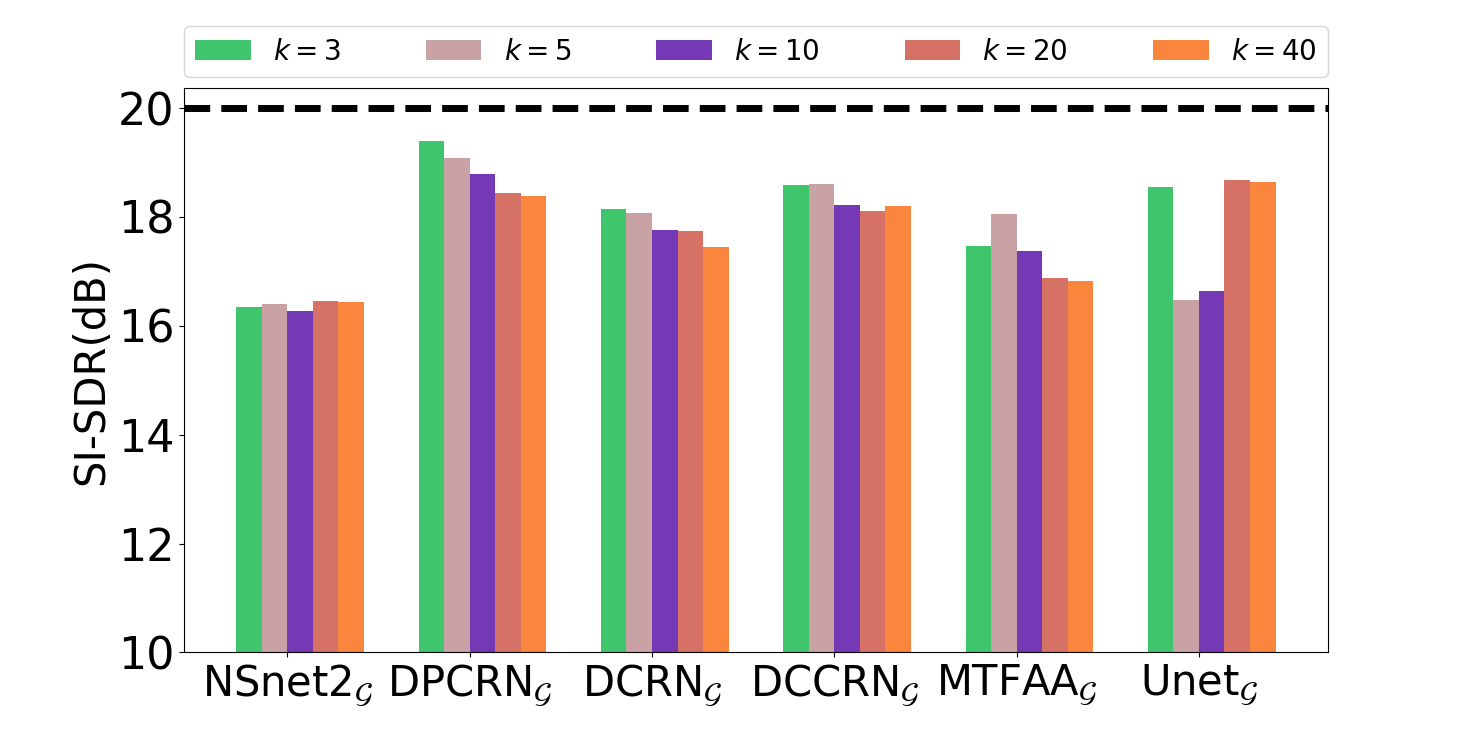}\vspace{-0.2cm}
        \centerline{\scriptsize (b) The average SI-SDR results}
    \end{minipage}
    \vspace{-0.45cm}
     \caption{The average WB-PESQ, SI-SDR of speech enhancement network-based GFT-SVD in different values of $k$.}
     \label{fig:kvalue}
\end{figure}

\vspace{-3pt}
Fig.~\ref{fig:kvalue} shows the influence of GFT-SVD based on ${{\textbf A}_{k}}$ in different cases of $k$ values. Note that $k$ represents the link number of a speech sample with neighbors.
We can clearly observe that when $k$ is equal to 3 or 5, the average PESQ and SI-SDR of the combination of GFT-SVD and DNN architectures can rank the best, While the performance of all models will decrease with the increased value of $k$. This is because GFT-SVD based on ${{\textbf A}_{k}}$ can capture strongly hidden information between neighbor speech sampling points, thus supporting more stable graph spectra of speech signals in the case of the graph frequency domain with $k=3,5$. 

\vspace{-12pt}
\subsection{Evaluation of Computational Efficiency}
\vspace{-5pt}
Table~\ref{table_par} reports the multiplier-accumulator operations (MACs) and network parameters ($\#$Param). 5 seconds of input audio are used to test the computational cost of the number of MACs per second. The last column results list the real-time factor (RTF) on Nvidia GeForce GTX 3090Ti. Observe from Table~\ref{table_par}, the combination of GFT-SVD and backbone networks performs better than the baseline with the same number of parameters and MACs. These results demonstrated the superiority of GFT-SVD in speech enhancement compared to the combination of STFT/GFT-EVD and backbone networks.

\begin{table}
	\caption{THE MACs AND PARAMETERS OF THE COMBINATION OF GFT-SVD AND DIFFERENT MODELS ON NO-REVERB DATASETS}
     \centering
    \small
    \def\arraystretch{0.91}
	\setlength{\tabcolsep}{8pt}
  \begin{tabular}{lcccc}
    \toprule[1.5pt] 
   \bf{Neural} & \bf Fourier  & \bf{MACs} & \#\bf{Param}& \multirow{2}*{\bf{RFT}}\\
   \bf{Network} & \bf Transform &(G/s) & (M)& \\
   \midrule \midrule
        NSnet2~\cite{XiaBRDCT20}  & \multirow{1}*{STFT} &{1.91}&{3.04}& {0.009}\\
         \textbf{NSnet2$_{\cal {G}}$}&GFT-SVD &1.92&3.04& 0.006\\\midrule
		DPCRN~\cite{abs-2107-05429}& STFT&{62.29}&{0.88}& {0.011}\\
  \textbf{DPCRN$_{\cal {G}}$}&GFT-SVD&62.49&0.88&0.005  \\ \midrule
		DCRN~\cite{PandeyLWS21}& STFT& {27.54}&{2.03}& {0.009} \\
  \textbf{DCRN$_{\cal {G}}$}&GFT-SVD&27.63&2.03&{0.003}  \\ \midrule
		DCCRN~\cite{HuLLXZFWZX20}& STFT&{109.53}& {4.33}& {0.031}\\
  \textbf{DCCRN$_{\cal {G}}$}&GFT-SVD&109.88 &4.33 &0.013 \\\midrule
	  MTFAA~\cite{ZhangWYW22}& STFT& {24.22}&{2.19}& {0.040}\\
   	\textbf{MTFAA$_{\cal {G}}$}&GFT-SVD&24.30 & 2.19 &0.012 \\\midrule
        BSRNN~\cite{bsrnn}& STFT& {3797}&{200}& {0.247}  \\
         \textbf{BSRNN$_{\cal {G}}$}&GFT-SVD&{3812}&{200}& {0.118}  \\ \midrule
            UNet~\cite{DBLP:conf/miccai/RonnebergerFB15}&STFT&37.47 &18.10&{0.019} \\ 
           G-UNet~\cite{ZhangP22}&GFT-EVD&32.23 & 18.10&0.025  \\     
            \textbf{UNet$_{\cal {G}}$} &GFT-SVD&37.47 &18.10&{0.009} \\     
      \toprule[1.5pt]
  \end{tabular}
  \vspace{-1.5em}
  \label{table_par}
\end{table}

\vspace{-5pt}
\subsection{Masking Visualization}
\vspace{-4pt}

\begin{figure}[t]
 \centering
 \vspace{-2pt}
   \begin{minipage}[t]{0.4\linewidth}
         \includegraphics[width=1.3in, height=2.1cm]{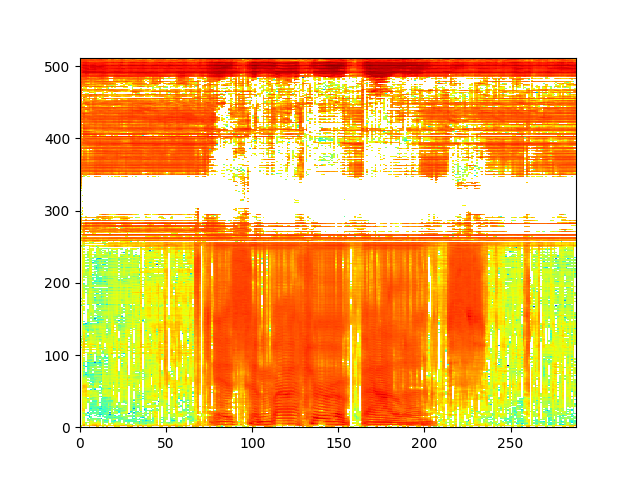}\vspace{-0.35cm}
        \centerline{\scriptsize (a) Mask.re (STFT)} \vspace{-10pt}
    \end{minipage}%
     \vspace{-2pt}
     \begin{minipage}[t]{0.4\linewidth}
    \includegraphics[width=1.3in,height=2.1cm]{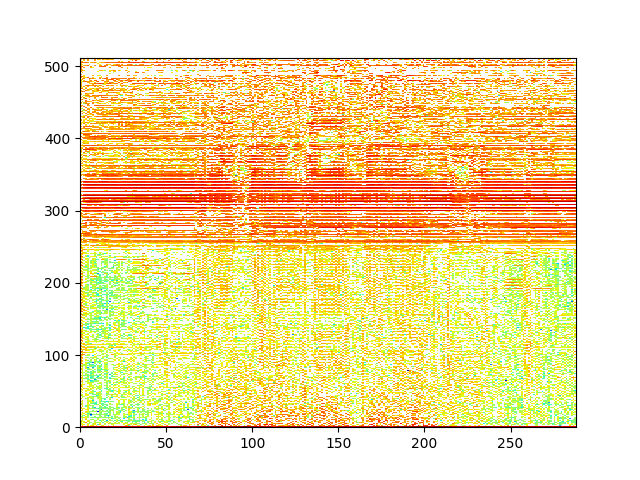}\vspace{-0.35cm}
         \centerline{\scriptsize(b) Mask.im (STFT)}  \vspace{-10pt}
         \end{minipage}
  \begin{minipage}[t]{0.4\linewidth}
    \includegraphics[width=1.3in,height=2.1cm]{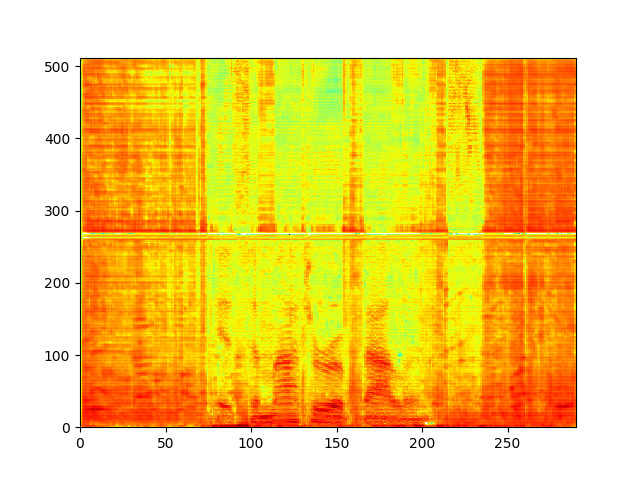}\vspace{-0.35cm}
        \centerline{\scriptsize (c) Mask.$1_{th}$ (GFT-SVD)} \vspace{-10pt}
         \end{minipage}
      \begin{minipage}[t]{0.4\linewidth}
    \includegraphics[width=1.3in,height=2.1cm]{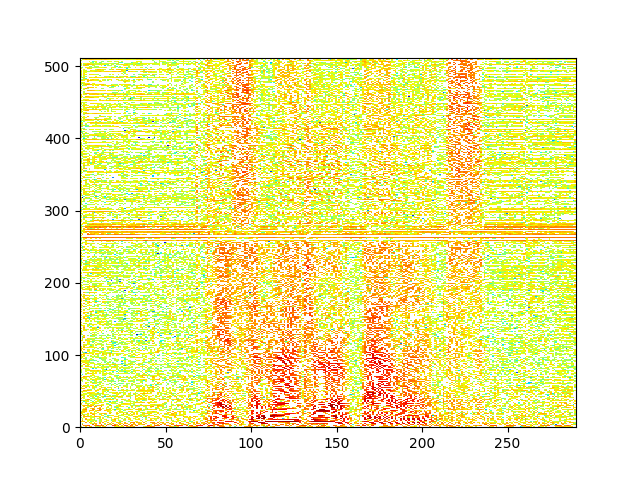}\vspace{-0.35cm}
        \centerline{\scriptsize (d) Mask.$2_{th}$ (GFT-SVD)}  \vspace{-10pt} 
     \end{minipage}\vspace{-0.2cm}
     \caption{ (a), (b) The real and imaginary parts of the estimated mask in setting DPCRN after STFT. (c), (d) The estimated two stream network mask in setting DPCRN$_{\cal {G}}$ after GFT-SVD.}
     \label{fig:mask}
    \vspace{-10pt}
\end{figure}

Fig.~\ref{fig:mask} confirms the improvement through visualization. Here we visualize the estimated masks in setting DPCRN, {DPCRN$_{\cal {G}}$ respectively on the VCTK+DEMAND dataset. From the Fig.~\ref{fig:mask}, we can conclude that the real-valued time-graph representation with GFT-SVD contributes to the training mask compared to the traditional complex T-F spectrogram. This time-graph representation significantly improves the phase prediction, helps alignment modeling of amplitude and phase.
		\label{fig3b}} 	

\vspace{-13pt}
\section{Conclusion}
\label{sec:conclusion}
\vspace{-5pt}

The paper introduces GFT-SVD into neural speech enhancement, which provides a real-valued time-graph representation. It supports an ability of alignment modeling of amplitude and phase in mask-based single-channel speech networks, leading to improving speech quality. Comparison with state-of-the-art systems on both DNS-2020 and VCTK+DEMAND datasets demonstrated the superiority of GFT-SVD over GFT-EVD, STFT on DNN architectures of speech enhancement.

\begin{small}
\bibliographystyle{IEEEtran}
\bibliography{strings,refs}
\end{small}

\end{document}